\documentclass[%
reprint,            
groupedaddress,
amsmath,amssymb,
aps,
prl,
floatfix,
longbibliography]{revtex4-2}

\usepackage{graphicx}
\usepackage{dcolumn}
\usepackage{bm, bbm}
\usepackage{xcolor}
\usepackage{float}
\usepackage{comment}
\newcommand{\ket}{{\rangle}}
\newcommand{\bra}{{\langle}}
\newcommand{\kb}{{\bar{k}}}

\begin{document}

\title{Efficient Plane-Wave Approach to Generalized Kohn-Sham Density-Functional Theory of Solids with Mixed Deterministic/Stochastic Exchange}
\author{Tucker Allen}
\author{Barry Y. Li}
\author{Tim Duong}
\author{Kajsa Williams}
\author{Daniel Neuhauser} 
\email{dxn@ucla.edu}
\affiliation{Department of Chemistry and Biochemistry, University of California, Los Angeles, CA, 90095, USA}

\date{\today}

\begin{abstract}

 An efficient mixed deterministic/sparse-stochastic plane-wave approach is developed for periodic generalized-Kohn-Sham density functional theory band-structure calculations with extensive $k$-point sampling, for any hybrid-exchange density functional.  The method works for very large elementary cells over many $k$-points, and we benchmark it on covalently bonded solids and molecular crystals with non-bonded interactions, for systems of up to 33,000 atoms. Memory and CPU requirements scale quasi-linearly with the number of atoms.

\end{abstract}
\maketitle

Generalized Kohn-Sham density functional theory (GKS-DFT), i.e., DFT with hybrid exchange-correlation (XC) functionals that include exact exchange, is now a cornerstone of electronic structure methods as such functionals reduce the self-interaction error of local- and semi-local DFT for both solids and molecules.\cite{Becke1993exact,Becke1993mixing,Seidl_PRB_1995} These functionals, particularly screened hybrids that separate short- and long-range exchange,~\cite{savin1997} capture the fundamental physics in both classes of systems.\cite{janesko_screened_2009,gaprenorm_roi_2013,roi_tddft_2009} Tuned hybrids control the balance between short- and long-range exchange through a range-separation parameter, $\gamma$, which is commonly obtained through either empirical fitting or first-principles calculation, e.g., by enforcing Koopmans' theorem.~\cite{baer_tuned_2010} For solids, short-range screened hybrids such as HSE06 produce lattice constants in good agreement with experiment for both semiconductors and insulators, while also yielding reliable bandgaps for semiconductors.\cite{heyd_hybrid_2003,heyd_energy_2005,tran_importance_2017,gianmarco_PRM} Another commonly used functional, the PBE0 global hybrid, uses a fixed fraction of exact exchange chosen based on perturbation theory arguments.\cite{perdew_generalized_1996,adamo_toward_1999} For molecules, long-range corrected (LC) hybrids enable proper description of charge-transfer and excitonic effects.\cite{savin1997,baer_density_2005} Additionally, dielectric-dependent hybrids have been successfully applied.\cite{skone_self-consistent_2014,skone_nonempirical_2016,zheng_effect_2017, bhandari_fundamental_2018, joo_solvation-mediated_2018, dunietz_2024}

Exact exchange scales usually quadratically with the number of $k$-points $N_k$. Many approaches have been developed to reduce the cost of evaluating exact exchange in both finite and extended systems.\cite{lin_adaptively_2016,wang_efficient_2020,wu2022,sharma_fast_2022,kokott_efficient_2024,Retting_thc_onk1,Bussy_real_onk1} However, efficient treatment of global and long-range hybrids that include a $G=0$ singular part of the exchange kernel remains a challenge as a larger $k$-point mesh is required to converge observables to the thermodynamic limit. 

In this article, we develop an efficient reciprocal-space plane-wave (PW) implementation of GKS-DFT. A cheap and accurate construction of the $k$-dependent exchange matrix is achieved for both small and very large unit cells. We introduce a general fitting procedure that uses the (semi)local-DFT wavefunctions sampled at the Brillouin zone center ($k=0$, the $\Gamma$-point) as a basis for the true $k$-dependent molecular orbitals (MOs) required for evaluating exact exchange. Then, to enable GKS-DFT calculations with dense $k$-point sampling we implement a mixed deterministic/sparse-stochastic approach, splitting the exchange kernel into low- and high-momentum components.\cite{bradbury_deterministicfragmented-stochastic_2023}

The starting point is a GKS-Hamiltonian
\begin{equation}
    H = K+V_{eN}+v_{H}+X+v_{xc},
\end{equation}
with the usual kinetic, nuclear, and Hartree terms; $X$ is the Fock operator under a general explicit exchange kernel $v$, 
\begin{equation}
    X(r,r')=-\rho(r,r') v(r-r'),
\end{equation}
and $v_{xc}(r)$ is a (semi)local-DFT XC potential for a short-ranged kernel $|r-r'|^{-1}-v(r-r')$. The explicit exchange kernel is usually made from short- and long-range parts, \cite{Becke1993exact,savin1997} 
\begin{equation}
\label{rsr}
    v(r-r') = \frac{\alpha+\beta \cdot \text{erf} (\gamma|r-r'|)}{|r-r'|}.
\end{equation}
The fraction of explicit exchange employed is dictated by the $\alpha$ and $\beta$ parameters, where at small interelectronic distance, i.e., $r\to r'$, $v(r-r')=\alpha|r-r'|^{-1}$, and at large distances, i.e., $|r-r'|\to \infty$, $v(r-r')=(\alpha+\beta)|r-r'|^{-1}$. Most standard hybrid functionals are available by appropriate combination of the $\alpha$, $\beta$, and $\gamma$ parameters. Various functionals are used here, including global Becke-type and range-separated hybrids that employ Fock exchange at short-range, long-range, or a mixture of both, see Table \ref{tab:functional_params}.

The sampling approach of the present work exploits the differing numerical weights of the long-range (low-$G$) and short-range (high-$G$) components of the overall Coulomb kernel, $|r-r'|^{-1}$. This is illustrated for the explicit exchange, Eq. \ref{rsr}, by working in Fourier space:
\begin{equation}
\label{vG}
    v(G)=\frac{4\pi}{G^2}(\alpha+\beta\cdot e^{-G^2/4\gamma^2}),
\end{equation}
where $G$ denotes a reciprocal lattice vector. The long-wavelength, low-$G$, contributions carry a large numerical weight and should be treated exactly, while the high-$G$ parts can become quite small and are amenable to data compression techniques.

The periodic near-gap formalism starts analogously to the non-periodic approach described in Refs.\cite{bradbury_deterministicfragmented-stochastic_2023, sereda_sparse-stochastic_2024}. Here, lower-case functions and coordinates refer to the supercells, and upper-case ones are used within a single unit (elementary) cell. The initial step is a cheap LDA (or a general DFT for local or semi-local functional) calculation for periodic systems on a supercell with $N_k$ unit cells, which yields $k$-space zero-order periodic MOs, labeled $|\Phi_{pk}\ket$. 

The global Bloch states, $\phi_{nk}(r)= \frac{1 }{\sqrt{N_k}} e^{ikr}\Phi_{nk}(r) $, are orthogonal on the supercell:
\begin{equation}
    \bra \phi_{nk} | \phi_{n'k'} \ket \equiv \int_{\rm supercell} 
    \phi_{nk}^*(r) \phi_{n'k'}(r) dr =\delta_{kk'} \delta_{nn'}.
\end{equation}
For the same $k$, the periodic part of the Bloch states is expanded in terms of plane-waves, $\Phi_{nk}(R)=\sum_G \Phi_{nk}(G)e^{iGR}$. These upper-case functions are orthogonal within a single unit cell, 
\begin{equation}
    \bra \Phi_{nk} | \Phi_{n'k} \ket \equiv \int_{\rm elem.cell} \Phi_{nk}^*(R) \Phi_{n'k}(R) dR =\delta_{nn'}.
\end{equation}

A ``band" of active orbitals near the Fermi level is then taken, and labeled as ``near-gap" states. Specifically, for each $k$-point we divide the states into several types: $N_{\rm core}$ (lower valence) states (which are of course above the inner-core states that are part of the norm-conserving pseudopotential (NCPP)); $N_v=N_{\rm occ}-N_{\rm core}$ upper valence; and $N_c$ low-lying conduction states.

The $M\equiv N_v+N_c$ near-gap orbitals are labeled as the MO-active space. Further, we introduce a subspace $A\subseteq M$, with $A\equiv A_v +A_c $, and $A_v (\le N_v )$ and $A_c(\le N_c)$ valence and conduction MOs, for which exchange is calculated explicitly. $A$ is labeled as the exchange-active space. The effect of the core states on the exchange is approximated as a perturbative scissor correction, discussed later.

The  GKS eigenstates on the supercell are then expanded in terms of zero-order MOs from the same $k$-point: 
\begin{equation}
    |\psi_{ik}\ket= \sum_j C^k_{ji}|\phi_{jk}\ket,
\end{equation}
where most integer indices extend over the $M$ active orbitals, with a similar $|\Psi_{ik}\ket= \sum_j C^k_{ji}|\Phi_{jk}\ket$ relation for a single unit cell.

For each $k$, $C^k$ is an eigenvector matrix of the $k$-dependent Hamiltonian matrix:
\begin{equation}
    H^k_{jl}\equiv\bra \phi_{jk} |  H   |\phi_{lk}\ket \equiv
    h^k_{jl} + X^k_{jl}, 
\end{equation}
where as usual,
\begin{equation}
 h^k_{jl} = \bra \Phi_{jk} |\frac {(k+\hat{G})^2 }{2} 
+V_{eN}+v_{H}+v_{xc}|\Phi_{lk} \ket.
\end{equation} 
Further
\begin{equation}
\label{xrr'}
    X(r,r')= -\sum_{m \kb} f_{m\kb} 
    \psi_{m \kb}(r)  v(r-r') \psi^*_{m \kb}(r'),  
\end{equation}
where the orbital occupations are introduced. The exchange matrix elements are then in real space:
\begin{equation}
\label{Xrealspace}
\begin{split}
&X^k_{jl} = \bra \phi_{jk} |  X   |\phi_{lk}\ket = -\sum_{m\kb} f_{m\kb}\iint dr dr' \\&
\phi_{jk}^*(r) \psi_{m\kb}(r) 
v(r-r') \psi^*_{m\kb}(r') \phi_{lk}(r'),
\end{split}
\end{equation} 
where the volume integrals extend over the supercell.

For the purpose of the exchange matrix elements only, we expand the $k$-dependent elementary functions in terms of the $\Gamma$-point functions (see also Ref.\cite{RL_2000})
\begin{equation}
\label{Bdef}
    |\Phi_{jk} \ket \simeq \sum_{j'} B^k_{j'j}|\Phi_{j'}\ket, 
\end{equation}
where $B^k_{j'j}=\bra \phi_{j'}|\phi_{jk}\ket$ and $|\Phi_{j'}\rangle \equiv |\Phi_{j',k=0}\rangle$. Note that not all $\phi_{jk}$ can be described properly, so for those that can’t be properly accounted for, i.e., those outside the exchange-active region, we write $B_{j'j}^k = \delta_{j'j}$. See SM for further details on the basis-set expansion. 

Thus, when used in the exchange part,
\begin{equation}
\label{Ddef}
    |\Psi_{jk} \ket \simeq \sum_{j'} D^k_{j'j}|\Phi_{j'}\ket, 
\end{equation}
where $D^k=C^k B^k$. Thus,
$X^k=\left(B^k \right)^\dagger Y^k B^k$, where the matrix element of $Y^k$ is 
\begin{equation}
\begin{split}
&Y^k_{jl} = 
- \frac{1}{N_k^2} \sum_{m \kb m' m''} f_{mk}  
D^\kb_{mm'} D^{\kb,*}_{mm''}\iint dr dr' \\&\Phi_{j}(r)\Phi_{m'}(r) e^{-i(k-\kb)(r-r')}
v(r-r') \Phi_{m''}(r') \Phi_{l}(r'),
\end{split}
\end{equation} where the $\Gamma$-point wavefunctions are real-valued. Upon Fourier transform, the exchange matrix in the reciprocal space reads:
\begin{equation}
\begin{split}
Y^k_{jl} = - \frac{1}{V_s} \sum_{m\kb m' m''} f_{mk}
D^\kb_{mm'} D^{\kb,*}_{mm''} \\
\sum_G \bra \Phi_{j}\Phi_{m'} |G\ket
     v(G+k-\kb)
    \bra G | \Phi_{m''}\Phi_{l} \ket,
\end{split}
\end{equation}    
where $V_s$ is the supercell volume. The momentum-space representation is readily shown to be:
\begin{equation}
    \label{Ylow}
    Y^{k}_{jl} 
    = - \sum_{i\kb G} 
    z^*_{j G i \kb} z_{l G i \kb } v(G-k+\kb),
\end{equation}
with rotated pair densities 
\begin{equation}
\label{auxvec}
    z_{l G i \kb }=\sqrt{\frac{f_{i\kb}}{V_s}} 
    \sum_t\bra G|\Phi_l \Phi_t\ket D^{\kb,*}_{ti},
\end{equation} 
and projections
\begin{equation}
\bra G | \Phi_{l}\Phi_{t} \ket\equiv 
\int_{\text{elem. cell}} \Phi_{l}(R) \Phi_{t}(R) e^{-iGR}dR, 
\end{equation}
and the exchange kernel is
\begin{equation}
\label{kernel}
    v(G+k-\kb) \equiv \int_{\rm supercell} v(r) e^{-i(G+k-\kb)r} dr.
\end{equation}

For $G\to0$, $v$ in Eq.(\ref{kernel}) could be singular, so we use a variant of the well-known Brillouin-supercell averaging.\cite{ismail-beigi_truncation_2006,arias_regularize2013} For small arguments, we replace $v(G-k+\kb)$ with a modified potential:
\begin{equation}
    \bar{v}(G-k+\kb) = \frac{\int \theta_b(p) v(G-k+\kb+p) dp}{\int \theta_b(p)dp},
\end{equation}
where $\theta_b(p)$ is the Brillouin theta function, for an orthorhombic lattice with a cell size of $L_x\cdot L_y \cdot L_z$: 
\begin{equation}
    \theta_b(p) = 
    \theta\left( \frac{\pi}{L_x}-|p_x|\right) 
    \theta\left( \frac{\pi}{L_y}-|p_y|\right) 
    \theta\left( \frac{\pi}{L_z}-|p_z|\right). 
\end{equation}
The integral is obtained via a Monte-Carlo procedure, for smaller $|G-k+\kb|$, more sampling points are used. For all simulations, 2 million Monte Carlo points are used, and we could easily use more points to reduce the stochastic error further.


Next, we split the summation over reciprocal lattice vectors to two parts, low and high: $Y_{pq}^k = Y^{L,k}_{pq} + Y^{H,k}_{pq}$. The numerical parameter separating low from high momenta, labeled $G_0$, is later varied to ensure convergence. For low-$G$, the summation is evaluated by explicitly applying Eq.(\ref{Ylow}) with $|G|<G_0$. For high $|G|$, we approximate
\begin{equation}
    \label{vhighG}
    v(G-k+\kb)\simeq v(G) \,\,\,\,\, |G|>G_0.
\end{equation}  
Then $Y^{H,k}_{jl} \simeq Y^{H}_{jl}$, i.e., independent of wavevector $k$. This can be written as:
\begin{equation}
     Y^{H}_{jl}
    \simeq-\frac{1}{V_s}
    \sum_{i,\kb,t,s,|G|> G_0}
    \bra\Phi_j \Phi_t|G\ket \bra G|\Phi_s \Phi_l\ket
    D^{\kb *}_{si} D^{\kb}_{ti} 
    v(G).
\end{equation} 
This approximation removes the need to Fourier transform $v(r)$ for all possible combinations of $(k,\kb)$ per Eq. \ref{kernel} in the high-$G$ space.


The next step is the fragmented-stochastic-exchange formulation of Ref.\cite{bradbury_deterministicfragmented-stochastic_2023},
\begin{equation}
    \label{FragStochX}
    \sum_G |G\ket v(G)\bra G|= \frac{1}{N_\xi}\sum_\xi |\xi\ket  \bra \xi|,
\end{equation}
where $\xi$ is a fragmented-stochastic basis, here made of a set of $N_\xi$ short random vectors in the high-${G}$ space. We define a projection $P(G)$ that randomly falls on a strip of the high-$G$ grid, which is randomly positive or negative within the strip and zero elsewhere. Combining the projector with the Coulomb kernel yields the basis vectors: $\bra G|\xi\ket = \pm  \sqrt{\frac{N}{L} v(G)}P(G)$, where $N$ is the length of the reciprocal-space grid being sampled and $L\sim \mathcal{O}(N/N_\xi)$ is the length of the random vector.  

The stochastic resolution of the identity above is formally exact only in the limit  $N_\xi \to \infty$, but in practice, the results converge rapidly, so $N_\xi \sim 500$ is generally sufficient. Eq.(\ref{FragStochX}) yields:
\begin{equation}
\label{Yhigh}
    Y^H_{jl} \simeq - \sum_{i\kb \xi} u^*_{j\xi i \kb} u_{l \xi i  \kb}
\end{equation}
with
\begin{equation}
    \label{zxi}
    u_{l \xi i  \kb} = 
    \sqrt{f_{i\kb}}
    \sum_t  \bra \xi | \phi_l \phi_t \ket D^{\kb,*}_{ti}.
\end{equation}

Note that $Y^L$ scales quadratically with the number of $k$-points, while $Y^H$ scales only linearly. It is therefore beneficial numerically to use a lower $G_0$,  so that only a few $G$-vectors contribute to $Y^L$. We show later that the value of $G_0$ can be quite small, so that most $G$-vectors can be represented stochastically with the $N_\xi$ auxiliary basis, which does not grow with elementary cell size or number of $k$-points.

Eqs.(\ref{Xrealspace})-(\ref{zxi}) give the complete expressions for the $X$ matrix. A technical point is that due to linear dependence considerations, the $B$ matrix is not square; only $\Phi_{jk}$ orbitals in the small exchange-active space $A$ are expanded, while their basis set, i.e., $\Phi_{j'}$ orbitals (Eq.\ref{Bdef}), encompasses the full active space $M$. For orbitals outside the exchange-active region, we could use a scissors-like expression,
\begin{equation}
\label{scissor}
    X^k_{jl} =\delta_{jl} X^k_{\bar{j} \bar{j}}, 
\end{equation}
for $j$ and $l$ in the lower $N_v-A_v$ space, where $\bar{j}$ is the lowest orbital in the exchange-active region $A$, and analogously for orbitals in the $N_c-A_c$ space. An alternative would be to include the contribution of orbitals outside the $A$ subspace stochastically.\cite{bradbury_bse_2023} 

\begin{figure}[H]
\centering
\includegraphics[width=3in]{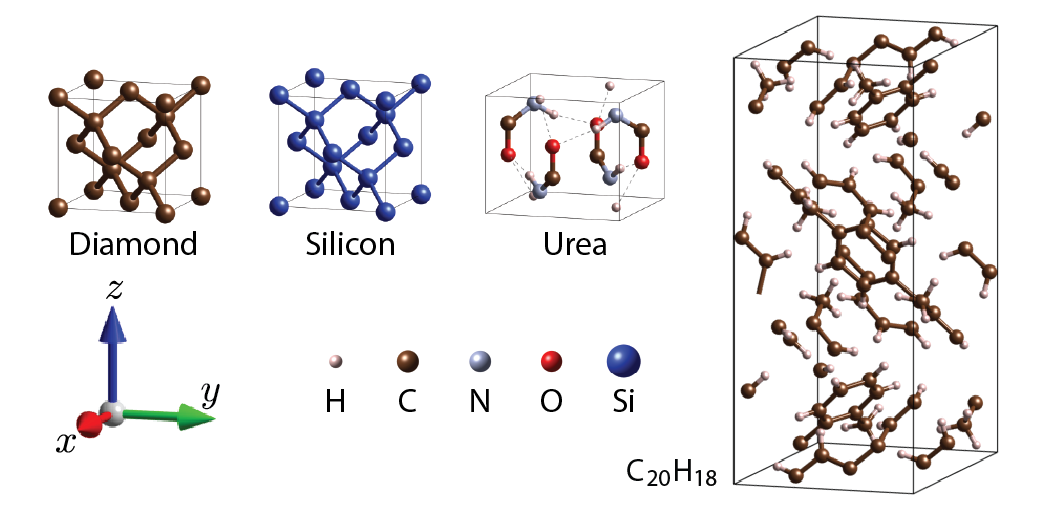}
\caption{\label{fig: units} Unit cells for systems studied.}
\end{figure}

The hybrid-exchange approach presented here is benchmarked on various orthorhombic lattices, including traditional covalently-bonded diamond and silicon (Si), and molecular crystals with $\pi$-$\pi$ interactions: urea and 1,4-Bis-(2-methyl-phenyl)-benzene (C$_{20}$H$_{18}$).\cite{diamond_si,urea,c20h18} 
The C$_{20}$H$_{18}$ unit cell consists of $\pi$-stacked layers of benzene rings arranged in a staggered geometry, see Fig.\ref{fig: units}. 

LDA-DFT calculations on uniform $k$-grids are performed using Troullier-Martins NCPPs and a kinetic-energy cutoff of 25 a.u$.$\cite{PerdewWang1992,TroullierMartins91} These LDA simulations provide $k$-dependent energies (converged to $10^{-8}$ a.u$.$) and one-electron wavefunctions $\Phi_{jk}$ that serve as the initial basis. To ensure basis-set convergence, the LDA-DFT calculations use at least five times more conduction than valence bands. All calculations are performed on standard 128-core AMD Rome processors. LDA simulations are parallelized over grid-points, while the present method is parallelized over both grid-points and wavefunctions.

\begin{table}[H]
\centering
\begin{ruledtabular}
\begin{tabular}{cccc}
\textbf{Functional} & $\alpha$ & $\beta$ & $\gamma$ $(\rm{Bohr^{-1}})$   \\ \hline
BNL             & 0            & 1            & 0.11  \\
CL (CAM-LDA0)   & 0.19         & 0.46         & 0.33  \\
HSE06           & 0.25         & -0.25        & 0.11  \\
PBE0            & 0.25         & 0            & 0 
\end{tabular}
\end{ruledtabular}
\caption{Hybrid exchange parameters per Eq.(\ref{rsr}).}
\label{tab:functional_params}
\end{table}

Electronic bandgaps, i.e., the difference between the conduction-band minimum (CBM) and valence-band maximum (VBM), are calculated with several hybrid functionals: BNL \cite{baer_density_2005}, CAM-LDA0 (CL) \cite{YANAI200451}, HSE06, and PBE0. Table \ref{tab:functional_params} shows the standard tabulated range-separation parameters used for CAM-LDA0 and HSE06. We use here the HSE06 value of $\gamma=0.11$ $\rm{Bohr^{-1}}$ for the BNL long-range hybrid. The choice of hybrid-functional will affect the rate of bandgap convergence with respect to the number of $k$-points. The BNL and CAM-LDA0 range-separated hybrids are long-ranged and exhibit slow convergence due to the $G\simeq0$ components of the exchange kernel. In contrast, HSE06 is short-ranged and exhibits a faster rate of convergence.

\begin{table}[H]
\centering
\begin{ruledtabular}
\begin{tabular}{c|cccc}
$k$-point Sampling & BNL & CL & HSE06 & PBE0 \\ \hline
$1 \times 1 \times 1$  & 3.73 & 4.87 & 1.77 & 2.50 \\
$2 \times 2 \times 2$  & 3.64 & 4.69 & 1.78 & 2.46 \\
$3 \times 3 \times 3$  & 3.88 & 4.58 & 1.84 & 2.57 \\
$4 \times 4 \times 4$  & 3.51 & 4.00 & 1.69 & 2.29 \\
$5 \times 5 \times 5$  & 3.79 & 4.18 & 1.74 & 2.48 \\
$6 \times 6 \times 6$  & 3.58 & 3.89 & 1.70 & 2.40 \\
$8 \times 8 \times 8$  & 3.61 & 3.78 & 1.73 & 2.44 \\
$10 \times 10 \times 10$ & 3.56 & 3.72 & 1.72 & 2.45 \\
\end{tabular}
\end{ruledtabular}
\caption{\label{si_data} Bandgaps (eV) for Si lattices as a function of the $k$-point grid size for several functionals.}
\end{table}

\begin{table}[H]
\centering
\begin{ruledtabular}
\begin{tabular}{c|cccc}
 $k$-point Sampling & BNL & CL & HSE06 & PBE0 \\ \hline
 $1 \times 1 \times 1$  & 6.68 & 7.65 & 4.75 & 5.49 \\
 $2 \times 2 \times 2$  & 6.31 & 6.90 & 4.37 & 5.04 \\
 $3 \times 3 \times 3$  & 6.13 & 6.52 & 4.18 & 4.81 \\
 $4 \times 4 \times 4$  & 5.91 & 6.38 & 4.15 & 4.75 \\
 $5 \times 5 \times 5$  & 5.86 & 6.45 & 4.30 & 4.86 \\
 $6 \times 6 \times 6$  & 5.74 & 6.32 & 4.22 & 4.78
\end{tabular}
\end{ruledtabular}
\caption{\label{c20h18_data} Bandgaps (eV) for $\text{C}_{20}\text{H}_{18}$ (152-atom unit cell) for various functionals as a function of the number of $k$-points.}
\end{table}

We first show results for small cell-size systems, specifically Si (additional data for diamond and urea is given in the SI \cite{SeeSIb}). For small systems, the exchange-active space $A$ includes all valence orbitals and a large number of conduction orbitals. The cutoff parameter $G_0$ is converged so that for all functionals, the bandgaps agree within $10$ meV with a fully deterministic calculation; $G_0=3$ a.u$.$ is found sufficient for Si and diamond, and $G_0=2$ a.u$.$ for urea. These are small values, so the number of $G$-vectors that need to be treated exactly per Eq.\ref{Ylow} is only $6.1$\%, $1.8$\%, and $1.7$\% of the respective $G$-spaces.     

\begin{figure}[H]
\centering
\includegraphics[width=3.5in]{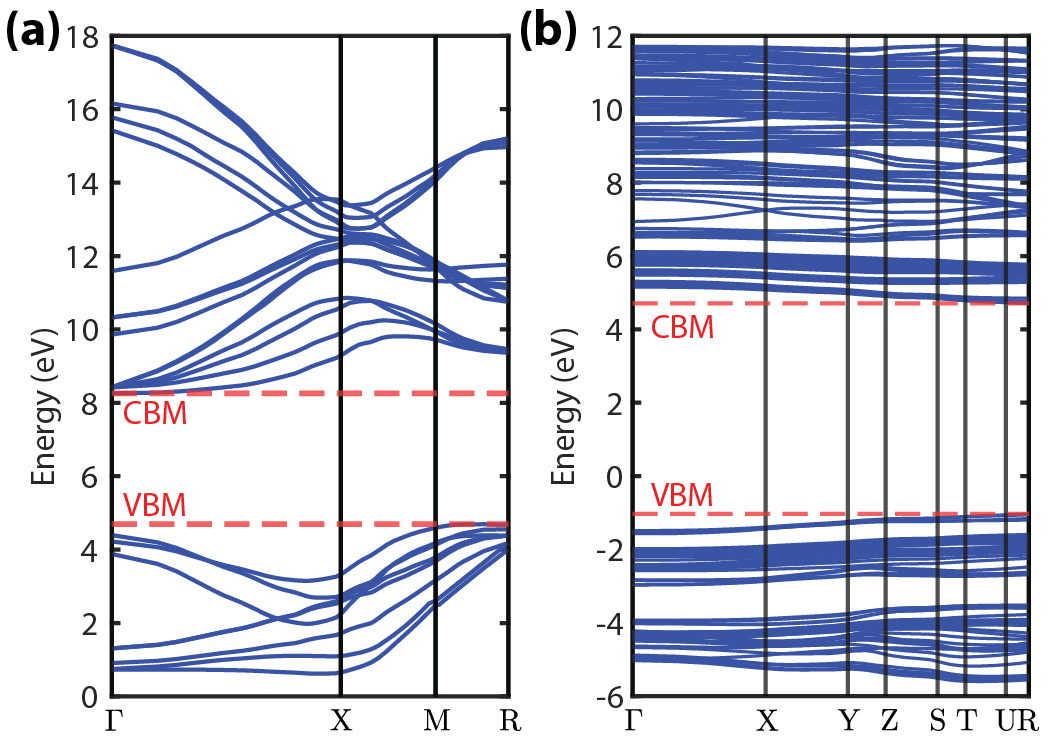}
\caption{\label{fig: si_bands} Band structures of \textbf{(a)} Si on a $10\times10\times10$ $k$-grid and \textbf{(b)} $\text{C}_{20}\text{H}_{18}$ on a $6\times6\times6$ $k$-grid via the BNL functional. Special symmetry points are based on an orthorhombic lattice.}
\end{figure}

Table \ref{si_data} shows bandgaps for Si, as well known, for global and long-range functionals, a large number of $k$-points is needed for convergence. Fig.\ref{fig: si_bands}(a) shows the band structure of Si on a  $ 10 \times 10 \times 10$ $k$-grid with the long-range BNL XC functional. Bandstructures for other systems as well as comparisons to traditional non-stochastic methods \cite{doi:10.1063/5.0005082} are provided in the SI (see also references \cite{Hoffman1987,HSE_JPCA_2022} therein).\cite{SeeSIc} 

We next move to a larger system, $\text{C}_{20}\text{H}_{18}$, with $152$ atoms within a single unit cell. Fig.\ref{fig: units} shows the unit cell, and Fig.\ref{fig: si_bands}(b) provides the band structure including 32,832 atoms using the BNL long-range hybrid. 
Table \ref{c20h18_data} provides bandgaps for various functionals as a function of the number of $k$-points. Due to the system's size, the molecular-orbital and exchange-active spaces are reduced to only include bands nearest to the Fermi level: $N_v=100$, $N_c=200$, $A_v=50$, and $A_c=100$. Selecting $N_v,A_v<N_{\text{occ}}$ (where $N_\text{occ}$ is the number of occupied valence bands) gives an error in the bandgap of roughly 200 meV, this could be remedied by a stochastic inclusion of the lower valence states, as in Ref.~\cite{bradbury_bse_2023}. The results are mostly insensitive to $A_c$, the number of exchange-active conduction bands, as long as a sufficiently large $N_c$ is used. Further, $G_0=1$ a.u$.$ here, so only $0.2 \%$ of the $G$-vectors are treated deterministically in the low-$G$ space while the remaining high-$G$ vectors are stochastically sampled with $N_\xi=$ 5000 sparse-stochastic vectors. 

\begin{figure}[H]
\centering
\includegraphics[width=3.5in]{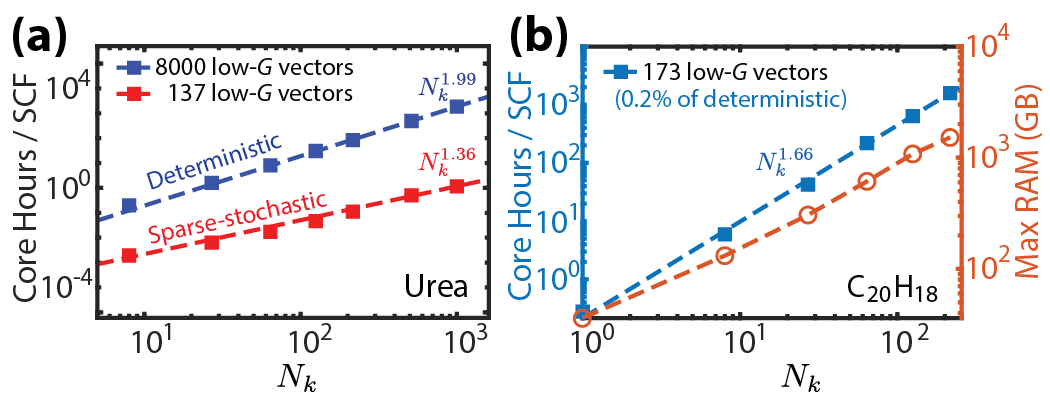}
\caption{\label{fig: comp_time} CPU core hours per SCF vs. number of $k$ points ($N_k$) on a logarithmic scale for \textbf{(a)} urea and \textbf{(b)} $\text{C}_{20}\text{H}_{18}$. For $\text{C}_{20}\text{H}_{18}$, maximum RAM usage is shown as a function of $N_k$.}
\end{figure}

We now move to discuss the computational cost of the new approach. Fig.\ref{fig: comp_time} (a) shows the CPU scaling with the number of $k$-points for urea using the PBE0 hybrid. The scaling with $N_k$ remains quadratic for the fully deterministic calculation (blue line), while the mixed deterministic/sparse-stochastic (red line) approach scales sub-quadratically with a much smaller prefactor. For example, the stochastic approach requires just 1 core hour per SCF iteration for a $10\times10\times10$ $k$-grid, whereas a deterministic calculation demands approximately 2000 core hours per SCF. 

Fig.\ref{fig: comp_time}(b) shows the effective scaling of the method for $\text{C}_{20}\text{H}_{18}$. In addition to CPU scaling, the maximum RAM required is provided. The linear scaling of RAM requirements with the number of $k$-points is substantially lower than a conventional PW implementation of general hybrid-exchange.

\begin{figure}[H]
\centering
\includegraphics[width=3.5in]{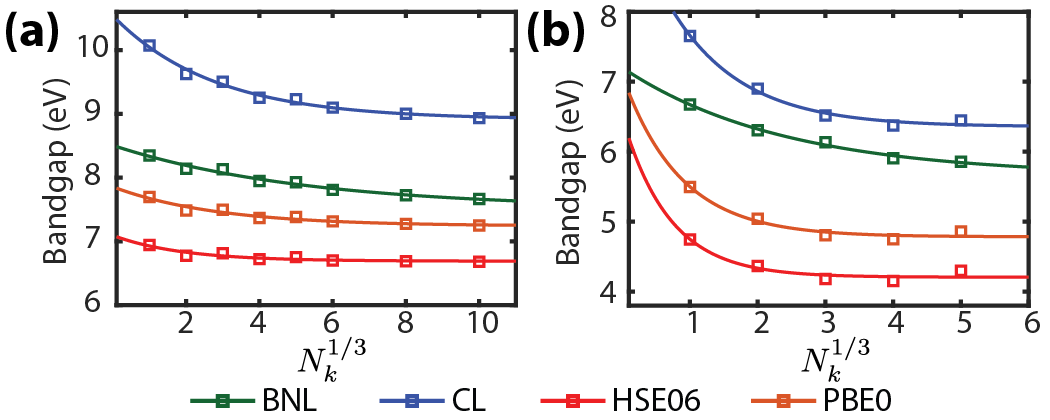}
\caption{\label{fig: expk} Single-exponential fits of bandgaps against the cubed root of the number of $k$-points ($N_k^{1/3}$) for \textbf{(a)} urea and \textbf{(b)} $\text{C}_{20}\text{H}_{18}$.}
\end{figure}

Fig.\ref{fig: expk} shows the exponential convergence of the bandgaps with the cubed root of the number of $k$-points ($N_k^{1/3}$) for urea and $\text{C}_{20}\text{H}_{18}$, respectively. This allows us to extrapolate to the thermodynamic limit, i.e., $N_k\to\infty$, with modest computational resources.   

The stochastic error associated with sampling the high-$G$ exchange (i.e., $Y^H$) was studied for urea using the CAM-LDA0 and PBE0 hybrid functionals. With $N_\xi = 5000$, the sample standard deviation of the $\Gamma$-point bandgap is small, below 10 meV. The error becomes less than 2 meV for bandgaps calculated on larger $k$-grids. The size of the sparse-stochastic basis could be reduced to even $N_\xi=500$, as the (tiny) stochastic error stems primarily from the Monte-Carlo sampling of the $G\simeq0$ parts of the exchange kernel.   

In conclusion, we developed and benchmarked a grid-based PW implementation of GKS-DFT for periodic systems. The method scales gently with $k$-points and provides significant speedups to deterministic calculations. This approach enables extensive $k$-point sampling for DFT with any hybrid-exchange functional, including long-range hybrids. The GKS-DFT energies converge with less than ten SCF iterations. This makes the present method appealing for band-structure calculations and post-DFT excited-state approaches that require GKS energies and wavefunctions as inputs. 

Future work will extend the static periodic GKS-DFT formalism to linear-response TDDFT for optical absorption spectra of solids. In the solid state, inclusion of a long-range Coulomb tail in the exchange kernel is required to produce excitonic effects and spectra in good agreement with experiment.~\cite{nanoquanta_2002, PRM_2023} In addition, this approach would be used to solve the GW-Bethe-Salpeter equation in extended systems where extensive $k$-point sampling is essential for accurate prediction of exciton binding energies.~\cite{PRB_2023}

\begin{acknowledgments}
    $Acknowledgments-$Useful discussions with Vojtech Vlcek and Laura Gagliardi are gratefully acknowledged.
    This work was supported by the U.S. Department of Energy, Office of Science, Office of Advanced Scientific Computing Research, Scientific Discovery through Advanced Computing (SciDAC) program under award number DE-SC0022198. Computational resources for simulations were provided by both the Expanse cluster at San Diego Supercomputer Center through allocations CHE240067, CHE240183, PHY240131, and PHY250047, under the Advanced Cyberinfrastructure Coordination Ecosystem: Services \& Support (ACCESS) program and the resources of the National Energy Research Scientific Computing Center, a DOE Office of Science User Facility supported by the Office of Science of the U.S. Department of Energy under contract no$.$ DE-AC02-05CH11231 using NERSC award BES-ERCAP0029462.
\end{acknowledgments}

\bibliography{apssamp}
\end{document}